\documentclass[MSNbibl,number,dvips]{arxstspdf}
\usepackage{flushend}
\usepackage{stfloats}

\volume{27}
\issue{2}
\pubyear{2012}
\firstpage{193}
\lastpage{196}
\doi{10.1214/12-STS376C} 
\referstodoi{10.1214/11-STS376}

\makeatletter
\def\eqref#1{(\ref{#1})}
\makeatother

\begin{document}
\begin{frontmatter}
\vspace*{6pt}
\title{Nonparametric Inference for Max-Stable~Dependence}%
\runtitle{Nonparametric Inference for Max-Stable Dependence}

\begin{aug}
\author[a]{\fnms{Johan} \snm{Segers}\corref{}\ead[label=e1]{johan.segers@uclouvain.be}}
\runauthor{J. Segers}

\address[a]{Johan Segers is Professor, Institut de statistique,
biostatistique et sciences actuarielles,
Universit\'e catholique de Louvain, Voie du Roman Pays 20, B-1348
Louvain-la-Neuve, Belgium \printead{e1}. }

\end{aug}



\end{frontmatter}

The choice for parametric techniques in the discussion article is
motivated by the claim that for multivariate extreme-value
distributions, ``owing to the curse of dimensionality, nonparametric
estimation has essentially been confined to the bivariate case''
(Section 2.3). Thanks to recent developments, this is no longer true if
data take the form of multivariate maxima, as is the case in the
article. A wide range of nonparametric, rank-based estimators and tests
are nowadays available for extreme-value copulas. Since max-stable
processes have extreme-value copulas, these methods are applicable for
inference on max-stable processes too. The aim of this note is to make
the link between extreme-value copulas and max-stable processes
explicit and to review the existing nonparametric inference methods.

\section{Extreme-Value Copulas}
\label{Sevc}

Let the random variables $Y_1, \ldots, Y_D$ represent the maxima in a
given year of a spatial process (e.g.,\ rainfall) that is observed at a
finite number of sites, $x_1, \ldots, x_D$, in a region $\mathcal{X}$
in space $\mathbb{R}^p$ (typically, \mbox{$p = 2$}). Let $F_1, \ldots, F_D$
be the marginal cumulative distribution functions, assumed to be
continuous. In the article, these are assumed to be univariate
generalized extreme-value distributions, an assumption that will not be
needed here.

The random variables $U_d = F_d(Y_d)$ are uniformly distributed on the
interval $(0, 1)$ and the joint cumulative distribution function $C$ of
the vector $U_1, \ldots,\break U_D$ is the copula of the random vector $Y_1,
\ldots, Y_D$:
%
\begin{equation}
\label{eqcopula}
C(u_1, \ldots, u_D) = \operatorname{Pr}( U_1 \le u_1, \ldots, U_D
\le u_D )
\end{equation}
for $0 \le u_d \le1$. The requirement that the random vector $Y_1,
\ldots, Y_D$ is max-stable entails
%
\begin{equation}
\label{eqmaxstable}
C^m(u_1^{1/m}, \ldots, u_D^{1/m}) = C(u_1, \ldots, u_D)
\end{equation}
for all $m > 0$. In \cite{Pickands81}, it was shown that \eqref
{eqmaxstable} holds if, and only if,
%
\begin{equation}
\label{eqA2C}
C(u_1, \ldots, u_D) = \exp\{ - r   A(v_1, \ldots, v_D) \},
\end{equation}
where $r = - \sum_{d=1}^D \log u_d$ and $v_d = - r^{-1} \log u_d$. The
domain of the Pickands dependence function $A$ is the unit simplex,
${\mathcal{S}_D}= \{ v \in[0, 1]^D \dvtx \sum_d v_d = 1 \}$.\break A~necessary
and sufficient condition for a function~$A$ on ${\mathcal{S}_D}$ to be
a Pickands dependence function is that
\begin{eqnarray}
\label{eqM2A}
&&A(v_1, \ldots, v_D)\nonumber
\\[-8pt]\\[-8pt]
&&\quad = \int_{\mathcal{S}_D}\max(v_1 s_1, \ldots, v_D s_D)  \, \mathrm
{d}M(s_1, \ldots, s_D)\hspace*{-12pt}\nonumber
\end{eqnarray}
for a Borel measure $M$ on ${\mathcal{S}_D}$ verifying the constraints
$\int_{\mathcal{S}_D}s_d  \, \mathrm{d}M(s_1, \ldots, s_D) = 1$ for
all $d \in\{1, \ldots,\break D\}$. In particular, $A$ is convex and $\max
(v_1, \ldots, v_D) \le A(v_1, \ldots, v_D) \le v_1 + \cdots+ v_D$.
In dimension $D = 2$, these two properties completely characterize
Pickands dependence functions (but not if $D \ge3$).

\section{Max-Stable Models}
\label{Smaxstable}

The representation in \eqref{eqA2C}--\eqref{eqM2A} is valid for
general max-stable copulas and therefore also holds for the
finite-dimensional distributions of the max-stable\break processes considered
in Section 6 in the article. The purpose of this section is to make
this relation explicit.

Consider the simple max-stable process
%
\begin{equation}
\label{eqmaxstabproc}
\quad Z(x) = \max_{j \ge1} [ S_j \max\{ 0, W_j(x) \} ], \quad x \in
\mathbb{R}^p,
\end{equation}
where $\{ S_j \}_{j=1}^\infty$ are the points of a Poisson process on
$\mathbb{R}_+$ with rate $s^{-2}\, \mathrm{d}s$ and where $W_1, W_2,
\ldots$ are iid replicates of a stationary stochastic process $W$ on
$\mathbb{R}^p$, independent of the previous Poisson process, and such
that $\operatorname{E}[ W^+(x) ] = 1$, where we write $W^+(x) = \max
\{ 0, W(x) \}$. Particular cases of this\break model include the so-called
Smith model \cite{smith1990}, the Schlather model \cite
{schlather2002} and the Brown--Resnick mod\-el~\cite
{kabluchkoschlatherdehaan2009}.\vadjust{\goodbreak}

The stationary, marginal distribution of $Z(x)$ in~\eqref
{eqmaxstabproc} is unit-Fr\'echet and the joint distribution function
of the vector $Z(x_1), \ldots, Z(x_d)$ is given by
\begin{eqnarray*}
&&\operatorname{Pr}[ Z(x_1) \le z_1, \ldots, Z(x_D) \le z_D ]
 \\
&&\quad = \exp\Bigl[ - \mu\Bigl( \Bigl\{ (s, w) : \max_d \bigl(w^+(x_d) / z_d\bigr) >
1/s \Bigr\} \Bigr) \Bigr]
\end{eqnarray*}
for $z_d > 0$, where $\mu$ is the intensity measure of the Poisson
point process $\{ (S_j, W_j) \}_{j=1}^\infty$. A simple calculation
shows that
\begin{eqnarray*}
&&\operatorname{Pr}[ Z(x_1) \le z_1, \ldots, Z(x_D) \le z_D ]
\\
&&\quad= \exp\Bigl( - \operatorname{E}\Bigl[ \max_d \{ W^+(x_d) /
z_d \} \Bigr] \Bigr).
\end{eqnarray*}
As a consequence, the copula of $Z(x_1), \ldots, Z(x_D)$ is given by
the extreme-value copula with Pickands dependence function
%
\begin{equation}
\label{eqW2A}
A(v_1, \ldots, v_D)
= \operatorname{E}\Bigl[ \max_d \{ v_d   W^+(x_d) \} \Bigr]
\end{equation}
for $(v_1, \ldots, v_D) \in{\mathcal{S}_D}$. As illustrated by the
computations in \cite{dehaanpereira2006}, the integral arising in
\eqref{eqW2A} can rarely be calculated analytically, unless $D = 2$.

To recover the spectral measure $M$ in \eqref{eqM2A} from the
distribution of the stochastic process $W$, let $R = \sum_d W^+(x_d)$.
On the event $R > 0$, consider the random vector $(W^+(x_1), \ldots,
W^+(x_D)) / R$. Then
\begin{eqnarray*}
\label{eqW2M}
&&\mathrm{d}M(s_1, \ldots, s_D)
 \\
&&\quad =\operatorname{Pr}(R > 0)   \operatorname{E}[ R \mid\forall d\dvtx
W^+(x_d) / R = s_d ; R > 0 ]
 \\
&&\quad\quad{}\cdot\operatorname{Pr}[ \forall d\dvtx W^+(x_d) / R \in\mathrm{d}s_d \mid R >
0 ].
\end{eqnarray*}

\section{Estimation}
\label{Sestimation}

Nonparametric estimators of the Pickands dependence function are
surprisingly easy to construct and calculate. The starting point is the
simple fact that for $u \in[0, 1]$ and for $(v_1, \ldots, v_D) \in
{\mathcal{S}_D}$, the extreme-value copula $C$ with Pickands
dependence function~$A$ satisfies
%
\begin{equation}
\label{eqA2Cbis}
 C(u^{v_1}, \ldots, u^{v_D}) = u^{A(v_1, \ldots, v_D)},
\end{equation}
as can be verified from \eqref{eqA2C}. Using \eqref{eqA2Cbis}, the
function~$A$ can be recovered from the copula $C$ in various ways, for
instance, through integrals of the form
%
\begin{eqnarray}
\label{eqintC}
&&\int_0^1 f ( C(u^{v_1}, \ldots, u^{v_D}) )   g(u)
\,\mathrm{d}u\nonumber
\\
&&\quad = \int_0^1 f(u^\alpha)   g(u)  \, \mathrm{d}u,
\\
&& \hspace*{-9pt}\quad
\alpha= A(v_1, \ldots, v_D)\nonumber
\end{eqnarray}
for well-chosen functions $f$ and $g$. Plugging estimators for $C$ and
solving for $\alpha$ then yields estimators for $A$.

A natural estimator for $C$ is the empirical copula. Let $(Y_{i1},
\ldots, Y_{iD})$, for $i \in\{1, \ldots, n\}$, be an independent
random sample from a distribution with continuous margins and copula
$C$. The empirical copula is defined as
%
\begin{eqnarray}
\label{eqempcop}
\hspace*{12pt}&&C_n(u_1, \ldots, u_D)\nonumber
 \\[-4pt]\\[-12pt]
&&\quad = \frac{1}{n} \sum_{i=1}^n I \{ F_{n1}(Y_{i1}) \le u_1, \ldots,
F_{nD}(Y_{iD}) \le u_D \},\nonumber
\end{eqnarray}
where $F_{nd}$ is the (marginal) empirical distribution function of
$Y_{1d}, \ldots, Y_{nd}$. Being based on multivariate ranks, the
empirical copula is invariant under monotone transformations of the
data. The empirical copula goes back to the seminal paper by R\"
{u}schendorf \cite{ruschendorf1976} and has been studied and applied
intensively, such as recently in \cite{segers2011,T05,vanWel07}.

Inserting the empirical copula into \eqref{eqintC} and solving for
$A(v_1, \ldots, v_D)$ produces simple and (almost) \mbox{explicit}
estimators. Particular instances are the\break Pickands estimator \cite
{Deheuvels91,GS09,gudendorfsegers2011,pickands1989} and the Cap\'era\`
a--\break Foug\`eres--Genest estimator \cite{CFG97,GS09,gudendorfsegers2011}.
The bivariate versions of these estimators are special cases of the
weighted estimator in \cite{pengqianyang2011}. Minimum-distance
estimators are another instance of this technique~\cite{BDV11}.
Standard errors can be obtained via resampling \cite
{bucherdette2010,RS09} or via empirical likelihood \cite{pengqianyang2011}.

A drawback of the nonparametric estimators of $A$ is that they
typically do not produce valid Pickands dependence functions---remember
the representation in \eqref{eqM2A} that such functions must satisfy.
A way to overcome this issue is by projecting a possibly invalid pilot
estimator $A_n$ onto the family of Pickands dependence functions, yielding
%
\begin{equation}
\label{eqAproj}
A_n^{\mathrm{proj}} = \operatorname{\arg\min}\limits_{A \in
\mathcal{A}_D} \int_{\mathcal{S}_D}(A_n - A)^2,
\end{equation}
where $\mathcal{A}_D$ denotes the family of all Pickands dependence
functions in dimension $D$ and where the integral is with respect to
some measure on ${\mathcal{S}_D}$. In general, the
infinite-dimensional least-squares problem in \eqref{eqAproj} does not
admit an explicit solution. Approximate solutions can be obtained by
performing the minimization over the (finite-dimensional) class of
Pickands dependence functions with discrete spectral measures supported
on a given, finite grid \cite{FGS08,gudendorfsegers2011}.

Finally, note that a nonparametric estimator $A_n$ can be transformed
into a parametric one via mini\-mum-distance or projection techniques: in
\eqref{eqAproj}, replace~$\mathcal{A}_D$ by the parametric model of
interest. If the model happens to be specified via the point process
representation \eqref{eqmaxstabproc}, then this technique requires the
calculation of the Pickands dependence function $A$ via~\eqref{eqW2A}.

\section{Testing}
\label{Stesting}

Nonparametric methods are particularly suitable for hypothesis testing.
Of special interest are the hypothesis of max-stability in general and
the goodness of fit of a parametric model in particular. In most cases,
critical values are computed via resampling methods.

Even if the data at hand are vectors of component-wise maxima, it is a
good idea to test whether it is safe to assume that the underlying
distribution is max-stable, in particular, when the end-goal is to
perform prediction and/or extrapolation. For bivariate extreme-value
copulas, the first two moments of the random variable $W = C(U_1, U_2)
= F(Y_1, Y_2)$ happen to satisfy a particular linear relation. The
sample moments of the random variables $W_{n1}, \ldots,\break W_{nn}$
defined by
\[
W_{ni} = \frac{1}{n} \sum_{t=1}^n I(Y_{t1} \le Y_{i1}, Y_{t2} \le
Y_{i2})
\]
can therefore be converted to a test statistic for the null hypothesis
of max-stability \cite{BenGenNes09,GhoKhoRiv98}.

Another approach for testing max-stability is by comparing the
empirical copula $C_n$ in \eqref{eqempcop} with the extreme-value
copula that has a given estimator $A_n$ as its Pickands dependence
function. For the bivariate case, Cram\'er--von Mises tests based on
the Pickands and Cap\'era\`a--Foug\`eres--Genest estimators are
describ\-ed in \cite{BDV11,KY10}.

Finally, the adequacy of the hypothesis of max-stability can be tested
by directly exploiting the copula max-stability relation \eqref
{eqmaxstable} through a comparison of $C_n(u_1, \ldots, u_D)$ with
$C_n^m(u_1^{1/m}, \ldots, u_D^{1/m})$ for various values of $m > 0$.
Cram\'er--von Mises type test statistics turn out to be particularly
effective~\cite{kojadinovicsegersyan2011}.

The goodness of fit of a parametric model can be tested by comparing
the fitted parametric estimator for $A$ with a nonparametric one \cite
{GKNY11}. For max-stable models arising through the point process
representation in \eqref{eqmaxstabproc}, the function $A$ has to be
computed through the relation \eqref{eqW2A}. Shape constraints such as
exchangeability can be tested similarly~\cite{kojadinovicyan2012}.\looseness=1

\section{Conclusion}
\label{Sconclusion}

Nonparametric methods yield an attractive alternative inference method
for max-stable dependence. Estimators and test statistics of the
Pickands dependence function are (almost) explicit, even in the
general, multivariate case. Moreover, as the procedures are based upon
the ranks of the data only, the step of modeling the margins can be
skipped (which is not to be confused with the false statement that the
uncertainty on the margins has been eliminated altogether).
Many of the methods described in this contribution are implemented in
the R package \textsf{copula} \cite{kojadinovicyan2010copula}.

\section*{Acknowledgments}
Funding was provided by
IAP research network Grant P6/03 of the Belgian government (Belgian
Science Policy) and by ``Pro\-jet d'ac\-tions de
re\-cher\-che con\-cer\-t\'ees'' Number 07/12-002 of the Com\-mu\-nau\-t\'e fran\-\c{c}ai\-se de Bel\-gi\-que,
granted by the Aca\-d\'e\-mie uni\-ver\-si\-taire de Lou\-vain.

%

\end{document}